\documentclass[12pt]{article}
\usepackage{microtype}

\usepackage{amssymb}
\usepackage{ragged2e}

\usepackage{graphicx}

\usepackage{xcolor}

\usepackage{amsmath}

\usepackage{amssymb}

\usepackage{booktabs}

\usepackage[absolute,overlay]{textpos}

\usepackage{appendix}

	\usepackage{verbatim}

    \usepackage{mathrsfs}

    \usepackage{pdfpages}
\usepackage{capt-of} % Para adicionar títulos às inclusões de PDF

\usepackage{caption}
\usepackage[normalem]{ulem}

\usepackage{float}     % Para usar [H] e forçar a posição da imagem
\usepackage{subfig}    % Para usar subfiguras
\usepackage{placeins}

\usepackage{longtable}

\usepackage{hyperref}

\usepackage{amsmath} \usepackage{array}

\usepackage{multirow}
\usepackage{authblk}
\usepackage[margin=1in]{geometry}

\usepackage{nameref}

\title{{Quantum Creation of a FRW Universe: applying the Riesz fractional derivative}}
\author[1 \footnote{daniel.canedo@estudante.ufjf.br }]{D. L. Canedo}
\author[2 \footnote{pmoniz@ubi.pt}]{P. Moniz}
\author[1 \footnote{gilneto@fisica.ufjf.br}]{G. Oliveira-Neto}

\affil[1]{Departamento de Física, Instituto de Ciências Exatas, Universidade Federal de Juiz de Fora, Juiz de Fora, MG CEP 36036-330, Brazil}
\affil[2]{Departamento de Física, Centro de Matemática e Aplicações (CMA-UBI), Universidade da Beira Interior, Rua Marquês d’Ávila e Bolama, 6200 Covilhã, Portugal}

\date{March 19, 2025}

\begin{document}

\maketitle

\begin{center}
 \section*{Abstract} \justify
In this work, we apply fractional calculus to study quantum cosmology. Specifically, our Wheeler-DeWitt equation includes a FRW geometry, a radiation fluid, a positive cosmological constant, and an \textit{ad-hoc} potential; we employ the Riesz fractional derivative, which brings a parameter $\alpha$, where $1 < \alpha \leq 2$,  appearing explicitly in the mentioned equation. We investigate numerically the tunnelling probability for the Universe to emerge using a suitable WKB approximation.  Our findings are as follows. 
When we decrease the value for $\alpha$, the tunnelling probability also decreases, suggesting that if fractional features could be considered to ascertain among different early universe scenarios, then the value $\alpha=2$ (meaning strict locality and standard cosmology) would be the most likely. Finally, our results also allow for an interesting discussion between selecting values for $\Lambda$ (in a non-fractional conventional set-up) versus balancing, e.g., both $\Lambda$ and $\alpha$ in the fractional framework. Concretely, the probability transition in the former if, e.g., $\Lambda=0.7$,  is very close to the value computed if in the latter we employ instead, e.g., $\Lambda=1.5$ and $\alpha=1.9397961$.
\end{center}

\section{Introduction}

Since the introduction of the Wheeler-DeWitt equation \cite{Witt1}\cite{Witt2}, intending to describe the Universe as a quantum mechanical system, quantum cosmology (QC) has steadily developed and achieved significant results concerning the quest to understand the origin of the Universe. For seminal contributions and recent reviews, cf. references \cite{Julio}-\cite{observable}, the latter conveying a broad discussion between relational observables and yet to uncover cosmological signatures. The framework of QC includes audacious scenarios, namely 
the spontaneous creation from nothing  \cite{Criação espontânea 1}-\cite{Criação espontânea 8}: our time and space emerge through a potential barrier, the universe (we will find ourselves in and observe) tunnelling through with a finite size and free from the initial singularity.  This concept  was broadly embraced, conveying the study of  the probability of tunnelling of the Universe for different cosmological models; see for example in references  \cite{Paulo}-\cite{Gil5}. \newline

Given the context of the above paragraphs it is immensely tempting to explore QC further, trying new tools to understand the Universe's origin better. One such tool that recently gathered interest is fractional calculus \cite{Herrman}, an extension of traditional calculus that allows for the modelling of complex systems and thus may be applied to describe phenomena in classical and quantum cosmology, see, e.g.,  \cite{Rasouli}\cite{Gonzalez} for a broad sample of recently published contributions. \newline

Originating from G. l'Hôpital and G. W. Leibniz works,  then subsequently developed by mathematicians such as Abel, Liouville and Riemann,  an innovative mathematical tool was introduced,   allowing the order of derivatives and integrals to be non-integer. This tool endowed the development of  \textit{fractional calculus} and is currently applied in several branches of physics, such as field theory, particle physics and atomic and molecular physics. In addition, fractional calculus has also been used in engineering, biophysics and biomedicine, with encouraging results. For more information, details see in  \cite{Herrman},\cite{Hilfer}-\cite{Manuel}, e.g., and the many references therein. \newline

Contemporaneously, within quantum mechanics, fractional calculus has allowed a generalization of path integrals and 
the Schr\"{o}dinger equation, constituting fractional quantum mechanics (FQM).  It has also gathered interest, and quite a few publications have appeared in the literature, pointing to potential benefits and falsifiable tests; the interesting notwithstanding, there still exist significant open problems to address in FQM; please see in references  \cite{Laskin1}-\cite{Rabei} for more particulars. In particular, despite all the progress and success in FQM, there are still questions concerning the tunnelling effect. Let us be more concrete. 
In $\cite{Oliveira}$ and 
$\cite{Hasan}$, the authors studied quantum tunnelling through 
delta potential barriers and 
through a rectangular barrier, respectively, using a space\footnote{Tunnelling with fractional time derivatives (Caputo derivative)  can be found in 
\cite{China}.} fractional Schr\"{o}dinger equation. The former paper found an analytical expression for the tunnelling probability as a function of the fractional parameter: the transition probability increased as $\alpha$ decreased. In the latter paper, the other authors retrieved a numerical solution where the probability transition decreased as $\alpha$ decreased.  The scenarios were not so different (regarding the potential barrier that could be formulated), and fractional calculus generated nevertheless discrepant results regarding the behaviour in terms of how $\alpha$ varies.  \newline

The application of fractional calculus in either classical or quantum cosmology has also been of recent interest. Regarding the latter (usually designated as fractional quantum cosmology (FQC)), let us point to the following references \cite{Moniz1}-\cite{Tare}, among which plenty more are therein mentioned. For example, the focus ranged from black hole thermodynamics to deSitter cosmological wave function. Notably, FQC usually means taking (spatial) fractional derivatives. An application of interest is the Riesz derivative; it affects the kinetic term of the Hamiltonian. Overall, for derivatives with a non-integer (e.g., fractional) order, we can suitably modify the Wheeler-DeWitt equation, which governs the quantum state of the Universe, and aim to gain access and discuss fractional dimensions and, most importantly, non-local\footnote{In fact, the Riesz fractional derivative operates as a non-local operator unless \(\alpha = 2\).} effects \cite{Oliveira}-\cite{Moniz1}. 
We can also anticipate scale-dependent geometries, which can emerge subtly as mimicked quantum gravity effects. \newline

To be more precise, the Riesz fractional derivative is considered non-local because its definition involves a convolution between a function and a kernel that considers the function's values over a range of points, not just at a single point. 
In classical calculus, derivatives measure a function's local rate of change at a given point. In contrast, fractional derivatives, particularly the Riesz fractional derivative, involve a non-local interaction with the function, where the behaviour at a point depends on the values of the function at other (even distant) points.
Mathematically, the Riesz fractional derivative of order is often defined as
\[ D^\alpha f(x) = \frac{1}{2\pi} \int_{-\infty}^{\infty} \hat{f}(k) \, |k|^\alpha e^{ikx} \, dk; \]
there is a  Fourier transformation, a power-law factor that introduces the fractional differentiation order.
This non-local nature arises because the Fourier transform representation incorporates the global information of the function (i.e., the entire function in the frequency domain) rather than just the local information around a point. Consequently, the Riesz derivative responds to changes in the function's behaviour over a wider region, constituting a non-local derivative.
This contrasts with the classical derivative, which is strictly local and focuses only on the function's behaviour near the point of interest.  In the context of a fractional Wheeler-DeWitt equation, the behaviour of the wave function in a specific region is influenced by the conditions in that region and others, possibly the entire domain. Thus, a broader influence of quantum mechanical features on semiclassical domains can be substantial. Namely, within a WKB tunnelling discussion. \newline

 Subsequently, considering the contextualized presentation in the above paragraphs, our present work bears a twofold purpose. On the one hand, we know the seemingly distinct results in \cite{Hasan} and  \cite{Oliveira} regarding tunnelling in FQM. In particular,  \cite{Hasan} employs a realistic (albeit simplified) approximation for the potential barrier, being rectangular and having a finite width and height, allowing us to estimate probability transitions and other effects discussed. Also, as suggested in \cite{Hasan}, fractional parameters seem to alter the tunnelling dynamics. However, the authors in \cite{Hasan} suggest that broader generic potentials should be probed to ascertain the issue in discussion: would the probability of transition increase or decrease with $\alpha$ increasing or decreasing?  We herein use FQC and a broader, generic potential; no plane waves are thus used (contrary to \cite{Hasan}) to contribute to that query discussion.   On the other hand, we are widening the scope of \cite{GCM} using the fractional Riesz derivative. It is tempting to check if a variation of $\alpha$ can mimic any other alteration we could make in, e.g. $\Lambda$ or another parameter in \cite{GCM}, and within which range (even if limited) this may be possible. Moreover, by investigating  the fractional Riesz 
 derivative in FQC, we will be studying the tunnelling of FRW model (via a WKB approximation) for a generic potential as a function of the fractional parameter $\alpha$. Our potential barrier was constructed in \cite{GCM} and bears the curvature constant of the Universe $(k)$; the cosmological constant $(\Lambda)$ that plays the role of dark energy and a parameter associated with the magnitude of the \textit{ad-hoc} potential $(\sigma)$. Thus, we are 
  interested in comparing our results with the scenarios in either \cite{Hasan}, \cite{GCM}, contributing to a still scarce study area. \newline
 
With this aim, our paper is structured as follows. In section \ref{seção2} we import from \cite{GCM} the Wheeler-DeWitt 
 equation and then construct the fractional Schrödinger equation for our model. We will use an approximate WKB solution and determine the tunnelling probability (which we label as $TP_{WKB}$). In section \ref{seção3}, we present numerical  results of $TP_{WKB}$ for different selected  parameters (namely, $\alpha,\Lambda,k,\sigma$), comparing the $TP_{WKB}$ obtained  with  
 \cite{GCM}. Finally, section \ref{conclusão} presents our conclusions and discussions about our work.

\section{ Fractional WKB Tunneling Probabilites} \label{seção2}

We want to study how fractional calculus modifies the quantum tunnelling model studied in \cite{GCM}, where Oliveira et al. studied the classical and quantum dynamics of the universe with radiation, cosmological constant and an ad hoc potential for different curvatures. \newline

{Let us therefore briefly review. Starting from the FLRW metric, using the ADM formalism \cite{Misner} and the Schutz variational 
formalism
\cite{Schutz1}\cite{Schutz2}, the total Hamiltonian for the 
Universe in \cite{GCM} is written  as:

\begin{equation} \label{Hamiltoniana total}
    N\mathcal{H}= -\frac{p^{2}_{a}}{12} + p_{T} - 3ka^{2} + \Lambda a^{4} + V_{ah},
\end{equation}
where $p_{a}$ and $p_{T}$  are the canonically conjugated momenta to $a$ and $T$, respectively. Here we are using the natural unit system where $\hbar=c=k_{B}=8\pi G = 1$. The Schutz variational formalism is used to recover the covariance that is lost when using the ADM formalism by rewriting the fluid in terms of potentials, one of which is entropy. In this way, the variable $T$, associated with the radiation fluid, plays the role of time in this model. For more information see the ``Appendix: Hamiltonian for the radiation ﬂuid'' in \cite{GCM}.The parameter $\Lambda $ is the cosmological constant and $V_{ah}$ is the 
\textit{ad hoc} potential, introduced in \cite{Gil4} and written  as:

\begin{equation}
    V_{ah} = - \frac{\sigma^{2} a^{4}}{(a^{3} + 1)^{2}}
\end{equation}
where $\sigma$ is a dimensionless parameter associated to the magnitude of that potential. \newline

The creation of this universe was studied within a quantized model. Specifically, the total Hamiltonian (\ref{Hamiltoniana total}) is transformed into an operator using the usual commutation relations. Afterwards, we  introduce a wave function $\Psi$ that depends on the canonical variables. Imposing that the total Hamiltonian operator annihilates the wave function, we obtain the Wheeler-DeWitt equation:  
 
\begin{equation} \begin{matrix} \label{Wheeller-DeWitt 1}
    \hat{H} \Psi(a,T) = 0,\\
    
    \Bigg{(}\frac{1}{12} \frac{\partial^{2}}{\partial a^{2}} -i\frac{\partial}{\partial T} - 3ka^{2} + \Lambda a^{4} - \frac{\sigma^{2} a^{4}}{(a^{3} + 1)^{2}} \Bigg{)} \Psi(a,T) = 0.
    
    \end{matrix}
\end{equation}  
Equation (\ref{Wheeller-DeWitt 1}) can be re-cast  in the form of a time-dependent Schrödinger 
equation (for more information see Section 2 in \cite{Gil6}). Using a new variable $\tau = -T$ we have:  

\begin{equation} \label{Wheeller-DeWitt 2}
 \Bigg{(}\frac{1}{12} \frac{\partial^{2}}{\partial a^{2}}  - 3ka^{2} + \Lambda a^{4} - \frac{\sigma^{2} a^{4}}{(a^{3} + 1)^{2}} \Bigg{)} \Psi(a,\tau) = -i\frac{\partial}{\partial \tau}\Psi(a,\tau).
\end{equation}

\noindent The application of fractional calculus, using the Riesz derivative in the kinetic term of the Hamiltonian 
operator\footnote{For the standard textbook Schrödinger equation , we have 
 $    i\hbar \frac{\partial \Psi}{\partial t} = - \frac{\hbar^{2}}{2m} \Delta \Psi + V\Psi
$,
whereas the fractional QM imports that  $\frac{\hbar^{2}}{2m} \to D_{\alpha}(- \hbar^{2} \Delta)^{\frac{\alpha}{2}}$ and with $1 < \alpha \leq 2$ $\cite{Laskin2}$.
} is given by \cite{Overture}, \cite{Moniz1}:  

\begin{equation} 
\hat{H}_{\alpha}(p,r) = D_{\alpha}|p|^{\alpha} + V(r),
\end{equation}  
where $D_{\alpha}$ is a coefficient. Moreover, equation  (\ref{Wheeller-DeWitt 2})
can be written as:

\begin{equation} \label{Wheeller-DeWitt 2 Frac}
 \Bigg{(}D_{\alpha} { \Bigg{[}\frac{\partial^{2}}{\partial a^{2}}}\Bigg{]}^{\frac{\alpha}{2}}  - 3ka^{2} + \Lambda a^{4} - \frac{\sigma^{2} a^{4}}{(a^{3} + 1)^{2}} \Bigg{)} \Psi(a,\tau) = -i\frac{\partial}{\partial \tau}\Psi(a,\tau),
\end{equation}
where we take $\frac{1}{12} = D_{\alpha}$, when $\alpha = 2$. \newline

E. Capelas de Oliveira and Jayme Vaz Jr, in \cite{Oliveira}, solved the fractional Schr{\"o}dinger equation for the delta and double delta potential, taking into account that the fractional derivative is a nonlocal operator. M. Hasan and B.P. Mandal, in chapter 4 in \cite{Hasan}, studied the tunneling time in space fractional quantum mechanics for a time-independent fractional Schr{\"o}dinger equation, whose potential barrier was rectangular. Differently from what was proposed by these authors in their studies, we will obtain the tunneling probability by solving the Wheeler-DeWitt equation, in the form of a time-dependent Schr{\"o}dinger equation, for a generic potential barrier and for a non-flat wave function. For this, we will use the $WKB$ approximation. \newline

So let us propose a solution for (\ref{Wheeller-DeWitt 2 Frac}) as in \cite{GCM} in the following form

\begin{equation} \label{função de onda 1}
    \Psi(a,\tau) := \psi(a)e^{-iE\tau}
\end{equation}
where we choose $\psi(a) = A(a) e^{i\phi(a)}$, with $A(a)$ being the amplitude, $\phi(a)$ the phase, and $E$ in (\ref{função de onda 1}) is the energy associated to the radiation fluid of the Universe, following 
\cite{GCM}.

Substituting (\ref{função de onda 1}) into (\ref{Wheeller-DeWitt 2 Frac}), we obtain the following expression:

\begin{equation}  \label{Wheeller-DeWitt 3}
 \Bigg{(}D_{\alpha} { \Bigg{[}\frac{\partial^{2}}{\partial a^{2}}}\Bigg{]}^{\frac{\alpha}{2}}   - 3ka^{2} + \Lambda a^{4} - \frac{\sigma^{2} a^{4}}{(a^{3} + 1)^{2}} \Bigg{)} \psi(a)e^{-iE\tau} = -i\frac{\partial}{\partial \tau}\psi(a)e^{-iE\tau} ,
 \end{equation}
 and therefore,

\begin{equation} \label{potencial efetivo 2}
 \frac{\partial^{\alpha}\psi(a)}{\partial a^{\alpha}} + \frac{1}{D_{\alpha}} (E - V_{eff}(a)) \psi(a) = 0.
 \end{equation}
 This last equation can be rewritten as
 \begin{equation} \label{potencial efetivo 3}
     \frac{\partial^{\alpha}\psi(a)}{\partial a^{\alpha}} + \mathcal{\hat{K}}(a)^{\alpha}\psi(a) = 0,
 \end{equation}
where $\mathcal{\hat{K}}(a)\equiv \Big{(}\frac{1}{D_{\alpha}} (E - V_{eff})\Big{)}^{\frac{1}{\alpha}}$ and $V_{eff}(a)$ is the effective potential
\begin{equation} \label{potencial efetivo 2a}
V_{eff}(a) \equiv 3ka^{2} - \Lambda a^{4} + \frac{\sigma^{2} a^{4}}{(a^{3} + 1)^{2}},
\end{equation}

The general solution for equation (\ref{potencial efetivo 3}) is given by:

\begin{equation}\begin{matrix}
    \Psi(a) = \frac{C_{1}}{{\mathcal{\hat{K}}(a)}^{\frac{1}{\alpha}}}e^{\pm i\int \mathcal{\hat{K}}(a) da}, \quad  \quad E>V_{eff},\\
\\

  \Psi(a) = \frac{C_{2}}{{\mathcal{\kappa}(a)}^{\frac{1}{\alpha}}}e^{\pm \int \kappa(a) da}, \quad  \quad E<V_{eff} ,   
\end{matrix}    
\end{equation}
where $C_{1}$ and $C_{2}$ are constants and $\kappa(a)\equiv\Big{(}\frac{1}{D_{\alpha}} ( V_{eff} - E)\Big{)}^{\frac{1}{\alpha}}$. When $\alpha=2$, the standard solution in $\cite{Merzbacher}$ is recovered.

The $WKB$ tunneling probability is given by $\cite{Merzbacher}$:

\begin{equation} \label{PTwkb2}
     TP_{WKB} \equiv \frac{4}{(2\theta + \frac{1}{2\theta})^{2}}
\end{equation}
where 
$\theta$
    measures 
   the height and thickness of the barrier for a given energy \cite{Merzbacher}. For our model, we have:

\begin{equation} \label{theta}
    \theta = e^{\int_{a_{L}}^{a_{R}} \kappa(a) da}.
\end{equation}
Here, $a_{L}$ and $a_{R}$ are the turning points where the energy line crosses the potential barrier on the left and right sides, respectively. Substituting (\ref{theta}) into (\ref{PTwkb2}), we obtain the tunnelling probability for a wave function, encountering a high and wide barrier $(\theta \gg 1)$ of an effective potential:

\begin{equation} \label{PTwkb3}
     TP_{WKB} \equiv \frac{4}{\Big{(}2e^{\int_{a_{L}}^{a_{R}} \kappa(a) da} + \frac{1}{2e^{\int_{a_{L}}^{a_{R}} \kappa(a) da}}\Big{)}^{2}}
\end{equation}
where $\kappa(a) = \Bigg{(}{\frac{1}{D_{\alpha}}  \Big{(}3ka^{2} - \Lambda a^{4} + \frac{\sigma^{2} a^{4}}{(a^{3} + 1)^{2}} - E\Big{)}}\Bigg{)}^{\frac{1}{\alpha}}$.

\section{Results} \label{seção3}

With the Fractional $WKB$ Tunneling Probability, equation (\ref{PTwkb3}), we are now able to compute the tunnelling probabilities for the creation of the universe:  an application of a simplified version of fractional calculus, in other words, the Riesz derivative in QC. Due the equation (\ref{Wheeller-DeWitt 3}) does not present an analytical solution for our model, to solve the tunneling probabilities (\ref{PTwkb3}) we will use the numerical solution using Maple Software. \newline

Equation (\ref{PTwkb3}) has $5$ (five) parameters: $(1)$ the curvature $k$, $(2)$ the radiation energy $E$, $(3)$ the cosmological constant $\Lambda$, $(4)$ the \textit{ad hoc} potential parameter $\sigma$ and $(5)$ the FC 
$\alpha$; $D_{\alpha}$ is a constant. 
To investigate the tunnelling probability, we will consider in expression (\ref{PTwkb3}) a function of selected 
free parameters, which we  will vary as we compute numerically and plot them in the following subsections,  while fixing others. Let us be more concrete. 
To compare our results with the 
model $\cite{GCM}$, we will use the same values for the non-fractional parameters. These values will be reported in each case that will be studied in subsection \ref{3.2}.
After several numerical calculations  of the tunnelling probability (\ref{PTwkb3}) for different values of $\alpha$, with $1 < \alpha \leq 2$, the following behaviours were registered.

%\subsection{Tunneling Probability when $\alpha=2$} \label{seção 3.1}
\subsection{Tunneling Probability when \texorpdfstring{$\alpha = 2$}{alpha = 2}} \label{seção 3.1}

Here,  we call attention to the constant $D_{\alpha}$ present in the tunneling probability equation (\ref{PTwkb3}), before we begin the studies for the different parameters, fractional and non-fractional, of the model. This constant is dependent on the parameter $\alpha$ and its definition involves arbitrariness. However, when $\alpha = 2$ the result must be the constant of the standard model, without the fractional calculus. For fractional quantum mechanics, the constant $D_{\alpha} = \frac{1}{2m}$ is recovered when $\alpha = 2$ \cite{Laskin1}-\cite{Laskin3}.\newline

 In this way, defining $D_{\alpha} = (\frac{1}{12})^{\frac{\alpha}{2}}$
and substituting $\alpha =2$  in equation (\ref{PTwkb3}), we obtain the same equation and results for the tunnelling probability as in the model in \cite{GCM}. For more details see the \nameref{Constante}.

%\subsection{Tunneling Probability when $\alpha \neq 2$} \label{3.2}

\subsection{Tunneling Probability when \texorpdfstring{$\alpha \neq 2$}{alpha not 2} \label{3.2}}

We now investigate quantitatively how the fractional calculus features, namely the presence of the Riesz derivative, modify the quantum tunnelling probabilities. To do that, we will study expression (\ref{PTwkb3}) as a function of selected parameters, while keeping others fixed. We will employ the same values used for the parameters in \cite{GCM} to compare with the model herewith discussed.

\subsubsection{$TP_{WKB}$ as a function of $E$} \label{subsubsection E}

Concerning the variation for the energy $E$, we will fix $\Lambda=1.5$ and $\sigma=-50$ 
like in \cite{GCM}. All values for the energy $E$ are below $V_{eff} = 691.51$, which is the maximum value for the barrier.  We solved numerically the equation (\ref{PTwkb3}) for different values of $E$, of $\alpha $ where $1 < \alpha < 2$ and for different values of geometry $k$. We found that when $E$ and $\alpha$ increase, the tunnelling probabilities increase too. When we vary only the geometry $(k)$ keeping the other parameters fixed, we obtain that $TP_{WKB}$ is larger for the hyperbolic type spatial section. This behaviour can be found in Figure \ref{Variando E} below.

\begin{figure}[H]
\centering
         \begin{minipage}[t]{0.7\textwidth}            
            \includegraphics[width=\linewidth]{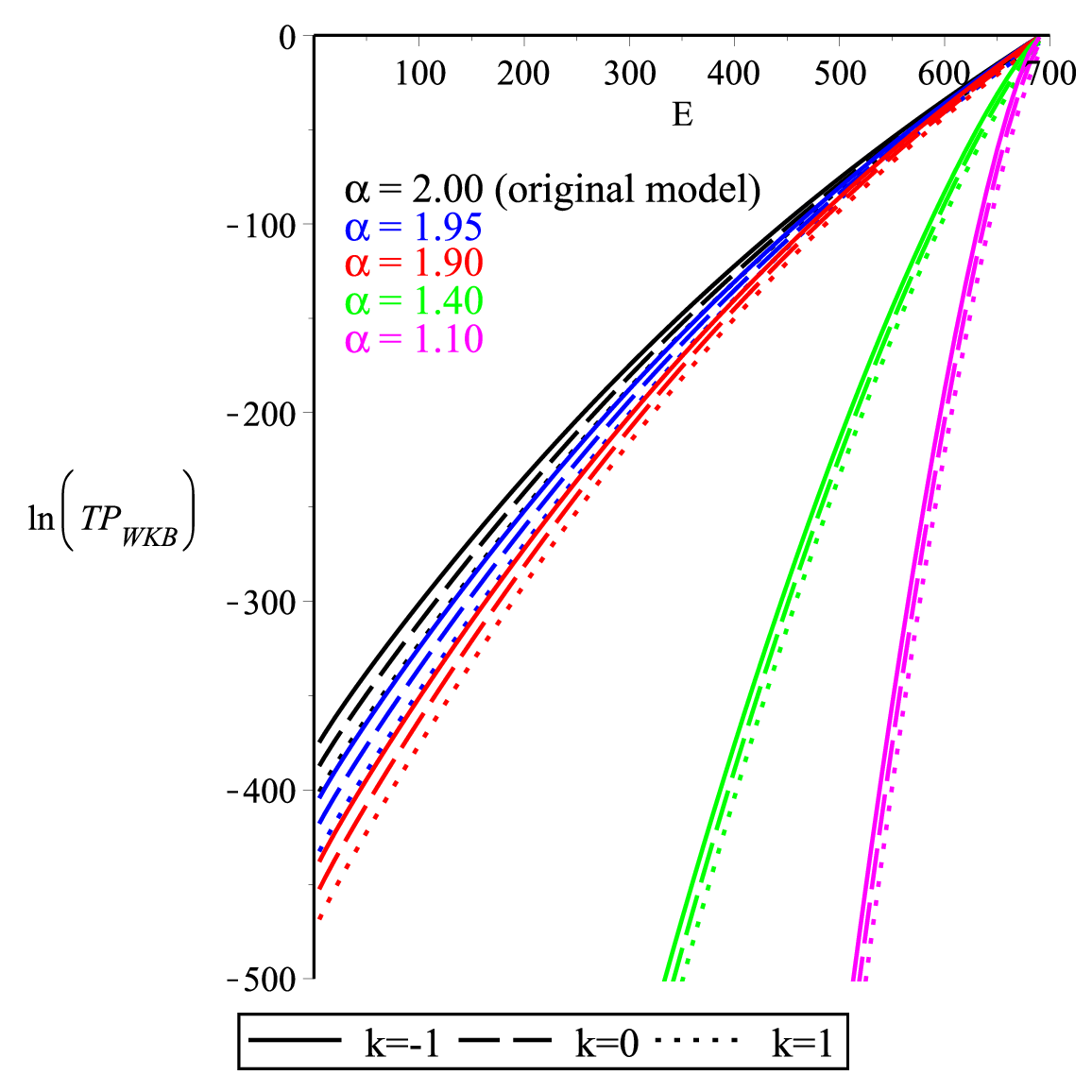}
            \caption{$TP_{WKB}$, in log scale, for the variation of the parameter $E$ and for different values of the parameters $\alpha$ and $k$, with $\Lambda = 1.5$ and $\sigma = -50$. The non-fractional parameters have the same value as in model \cite{GCM}. }
            \label{Variando E}
        \end{minipage}
    \end{figure}

    \FloatBarrier

\newpage
  
\subsubsection{$TP_{WKB}$ as a function of $\Lambda$}
The following study was for the cosmological constant $\Lambda$, playing the role of dark energy in this model. Solving 
numerically  equation (\ref{PTwkb3}) for different values of $\Lambda$ and $\alpha$, where $1 < \alpha < 2$, with $k=0,\pm1$, but with the non-fractional parameters $\sigma=-50$ and $E=690$ fixed,  we can see in Figure \ref{Variando L} that de $TP_{WKB}$ increase when $\alpha$ and $\Lambda$ increase and is larger for $k=-1$ geometry.
\begin{figure}[H]
\centering
         \begin{minipage}[t]{0.7\textwidth}            
            \includegraphics[width=\linewidth]{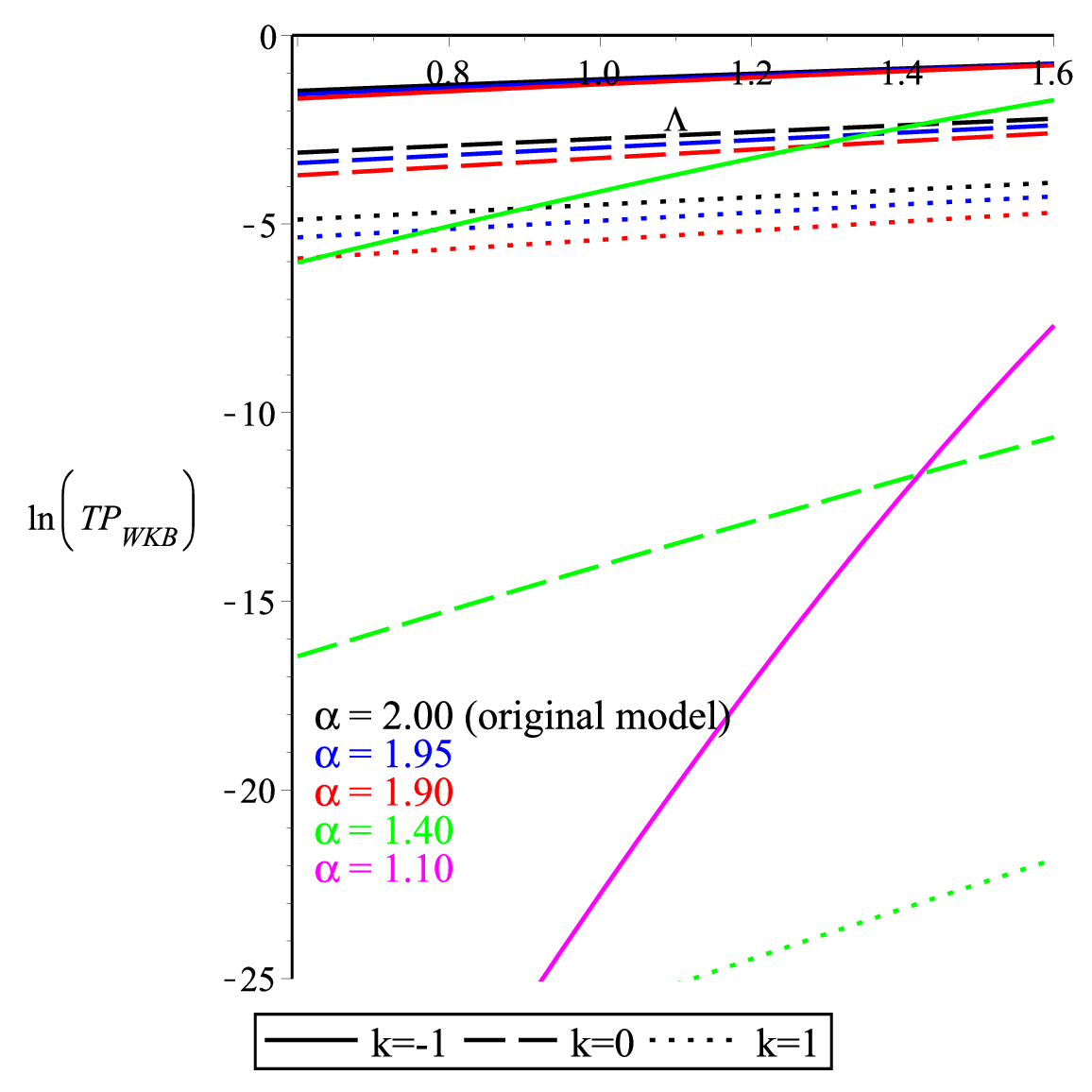}
            \caption{$TP_{WKB}$, in log scale, for the variation of the parameter $\Lambda$ and for different values of the parameters $\alpha$ and $k$, with $\sigma=-50$ and $E=690$ fixed. The non-fractional parameters have the same value as in  model \cite{GCM}. In this figure, we can only see the $k=-1$ line for $\alpha=1.10$. }
            \label{Variando L}
        \end{minipage}
    \end{figure}    
    \FloatBarrier

\newpage

\subsubsection{$TP_{WKB}$ as a function of $\sigma$}
Let us now vary the parameter $\sigma$, where $\sigma$ is a dimensionless parameter associated with the magnitude of the 
\textit{ad hoc} potential \cite{Gil4}. Following the same procedure, fixing the parameters $E=680$ and $\Lambda=1.5$, we vary and take several values of $\alpha$ and $\sigma$ for different space-like sections $(k)$. The 
results are in Figure \ref{Variando s}, and we can see that the $TP_{WKB}$ increase when $\sigma$ and $\alpha$ increase. The biggest value for the $TP_{WKB}$ is for $k=-1$ geometry if compare to $k=0,1$ geometries.
\begin{figure}[H]
\centering
         \begin{minipage}[t]{0.7\textwidth}            
            \includegraphics[width=\linewidth]{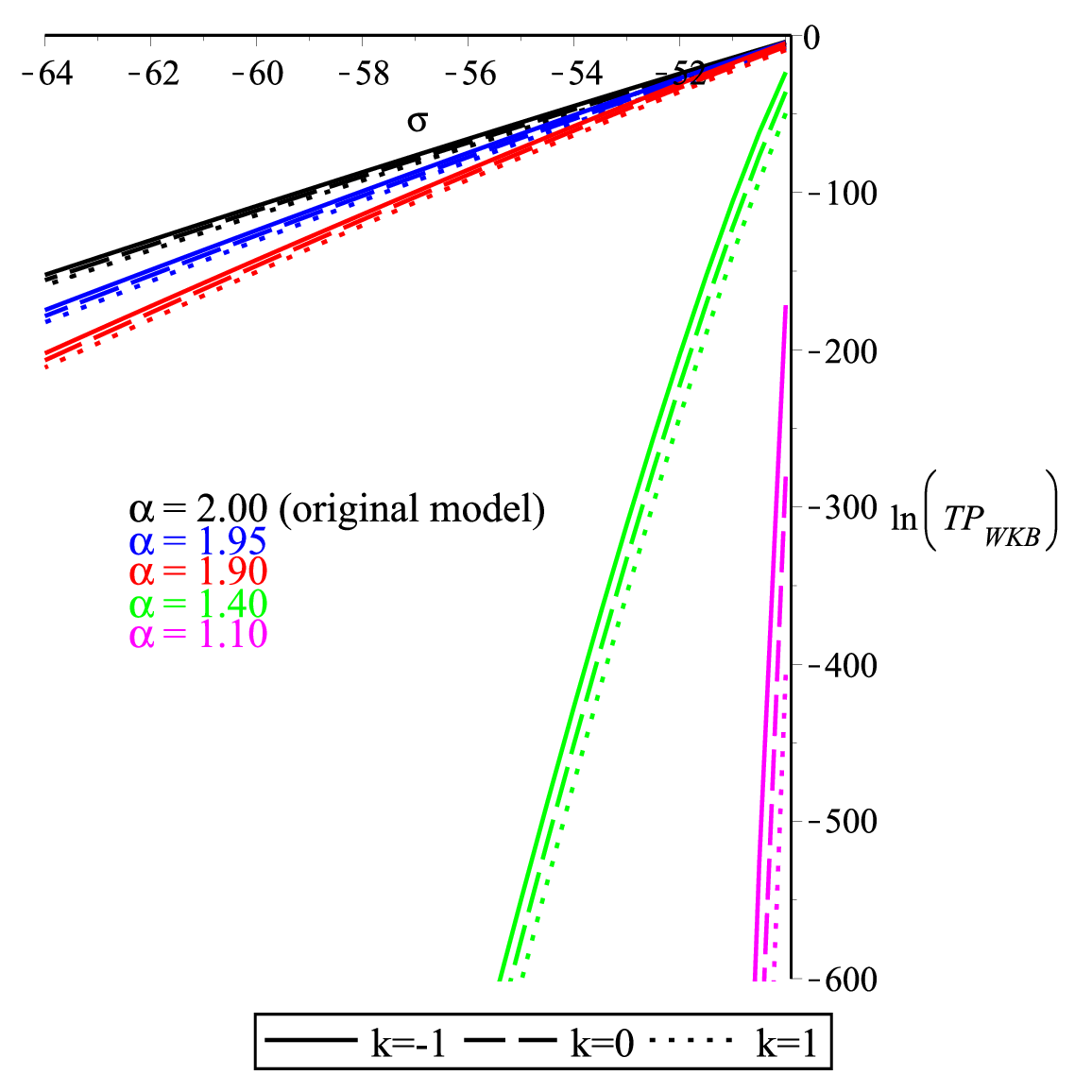}
            \caption{$TP_{WKB}$, in log scale, for the variation of the parameter $\sigma$ and for differents values of the parameters $\alpha$ and $k$, with $\Lambda=1.5$ and $E=680$ fixed. The non-fractional parameters have the same value as  the model \cite{GCM}.}
            \label{Variando s}
        \end{minipage}
    \end{figure}

    \FloatBarrier

\newpage

\subsection{Analysis and implication of the Riesz derivative in the calculation of the tunnelling probability.}

The Riesz derivative is an operator often used in fractional calculus and acts only on the kinetic term of the Hamiltonian operator (\ref{Wheeller-DeWitt 2 Frac}). Thus, the effective potential (\ref{potencial efetivo 2a}) is identical to the effective potential of the model in \cite{GCM}. In subsubsection \ref{subsubsection E}, we noticed that the $TP_{WKB}$ decreases when we decrease the value of the parameter $\alpha$, even when the energy $E$ remains the same as the value used in the model as in \cite{GCM}. In the non-fractional model, to decrease the probability of tunnelling while keeping the energy fixed, we need to modify the potential barrier to become e.g. wider and higher, whereas within the Riesz derivative discussion, this can go quite different. Let us be more specific. \newline

In figures \ref{V, k=-1}, \ref{V, k=0} and \ref{V, k=1} we put the potential barrier in three different cases: $(i)$ for the fractional case (blue line), $(ii)$ non-fractional with $\Lambda = 0.7$ (red line) and $(iii)$ the non-fractional model (green dot line) as in \cite{GCM}, i.e. the same $\Lambda, \sigma, k$ as in the blue line. The energy $E=200$ illustrates how the energy line intersects the barrier. We can see that the potential barrier in cases $(i)$ and $(iii)$ coincide;  however, they generate different $TP_{WKB}$ for the same energy, as per informed from figures 1-3.\newline

\begin{figure}[h]
  \centering
  \includegraphics[width=0.6\textwidth]{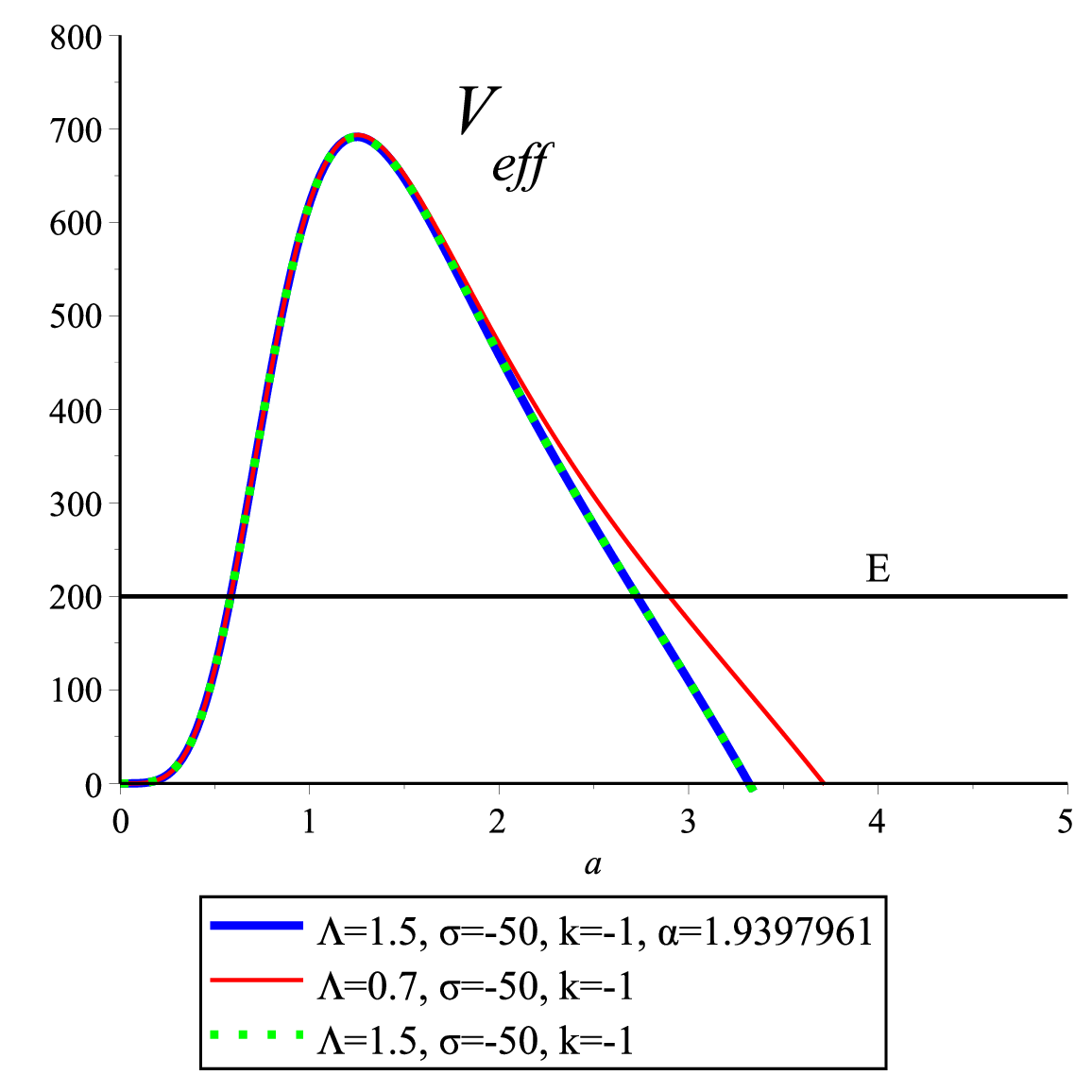}
  \caption{Effective potential for:  fractional (blue line), $\Lambda=0.7$ (red line) and  \cite{GCM} model (green dot line), for $k=-1$.}
  \label{V, k=-1}
\end{figure}

\begin{figure}[H]
  \centering
  \includegraphics[width=0.6\textwidth]{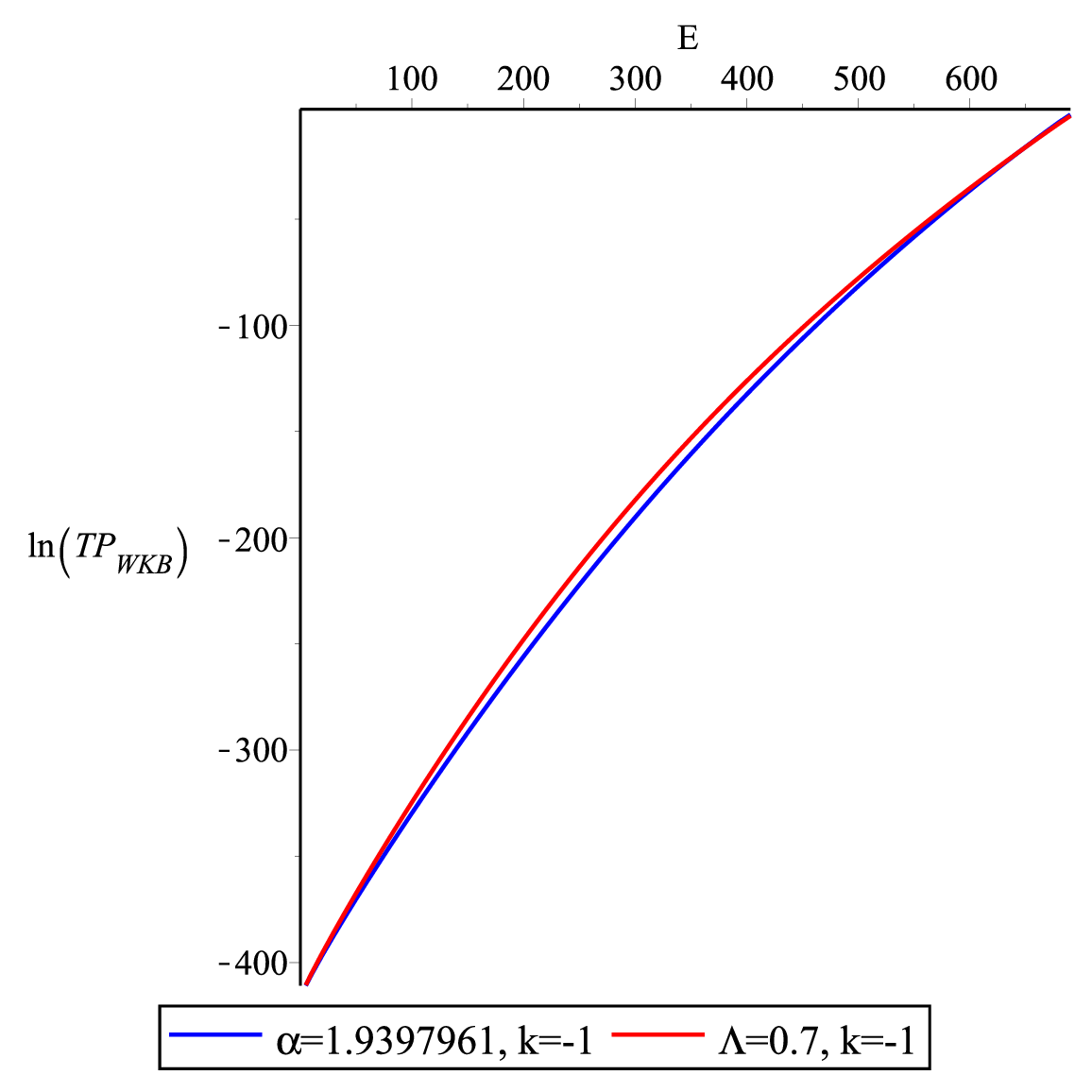}
  \caption{Comparison of $TP_{WKB}$, in logarithmic scale, for different energies $E$ in the fractional (blue line) and non-fractional $(\Lambda=0.7)$, red line, models for $k=-1$.}
  \label{F não-F k=-1}
\end{figure}
\newpage

\begin{figure}[H]
  \centering
  \includegraphics[width=0.55\textwidth]{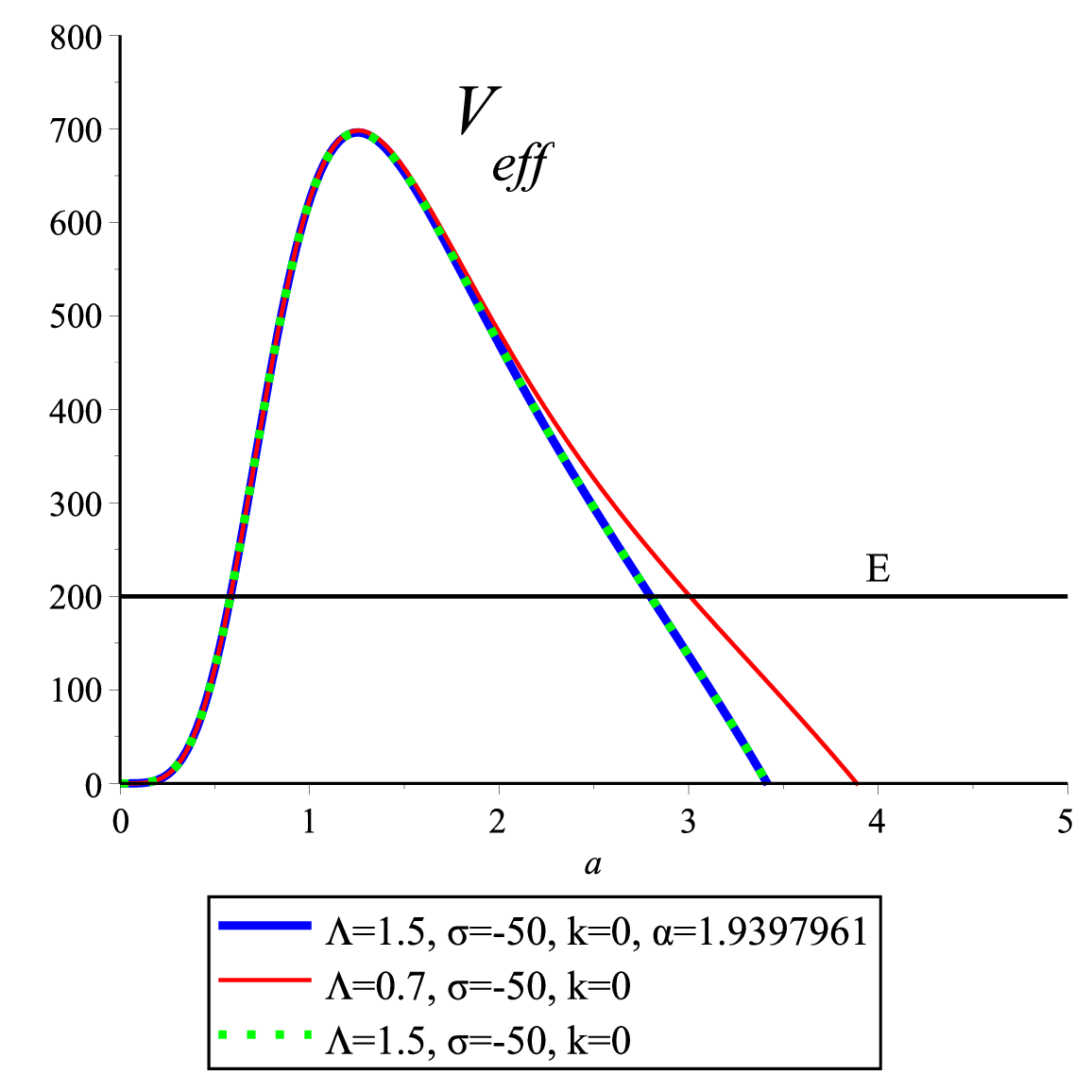}
  \caption{Effective potential for: fractional (blue line), $\Lambda=0.7$ (red line) and  \cite{GCM} model (green dot line), for $k=0$.}
  \label{V, k=0}
\end{figure}

\begin{figure}[H]
  \centering
  \includegraphics[width=0.55\textwidth]{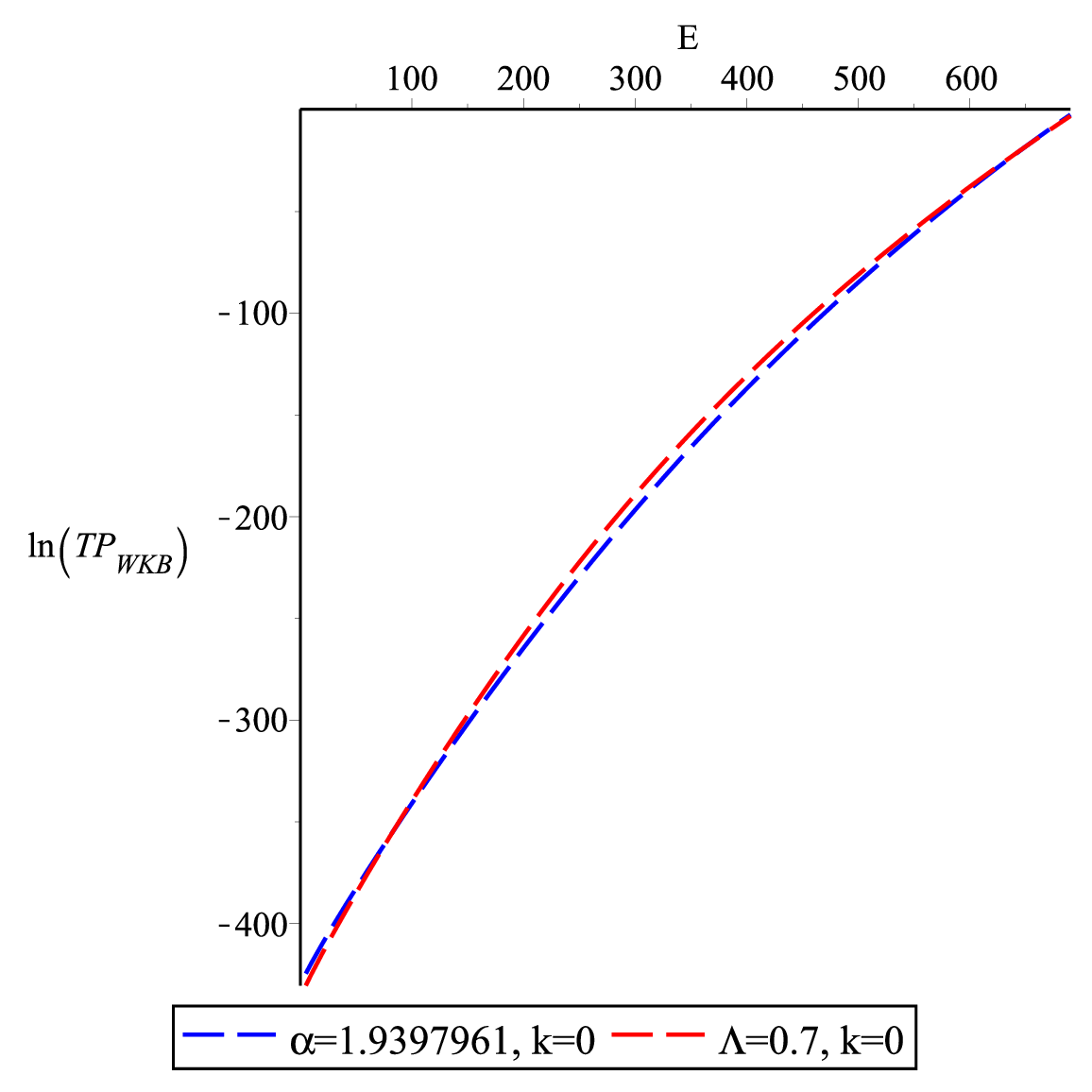}
  \caption{Comparison of $TP_{WKB}$, in logarithmic scale, for different energies $E$ in the fractional (blue slash line) and non-fractional $(\Lambda=0.7)$, red slash line, models for $k=0$.}
  \label{F não-F k=0}
\end{figure}

\FloatBarrier

\begin{figure}[H]
  \centering
  \includegraphics[width=0.55\textwidth]{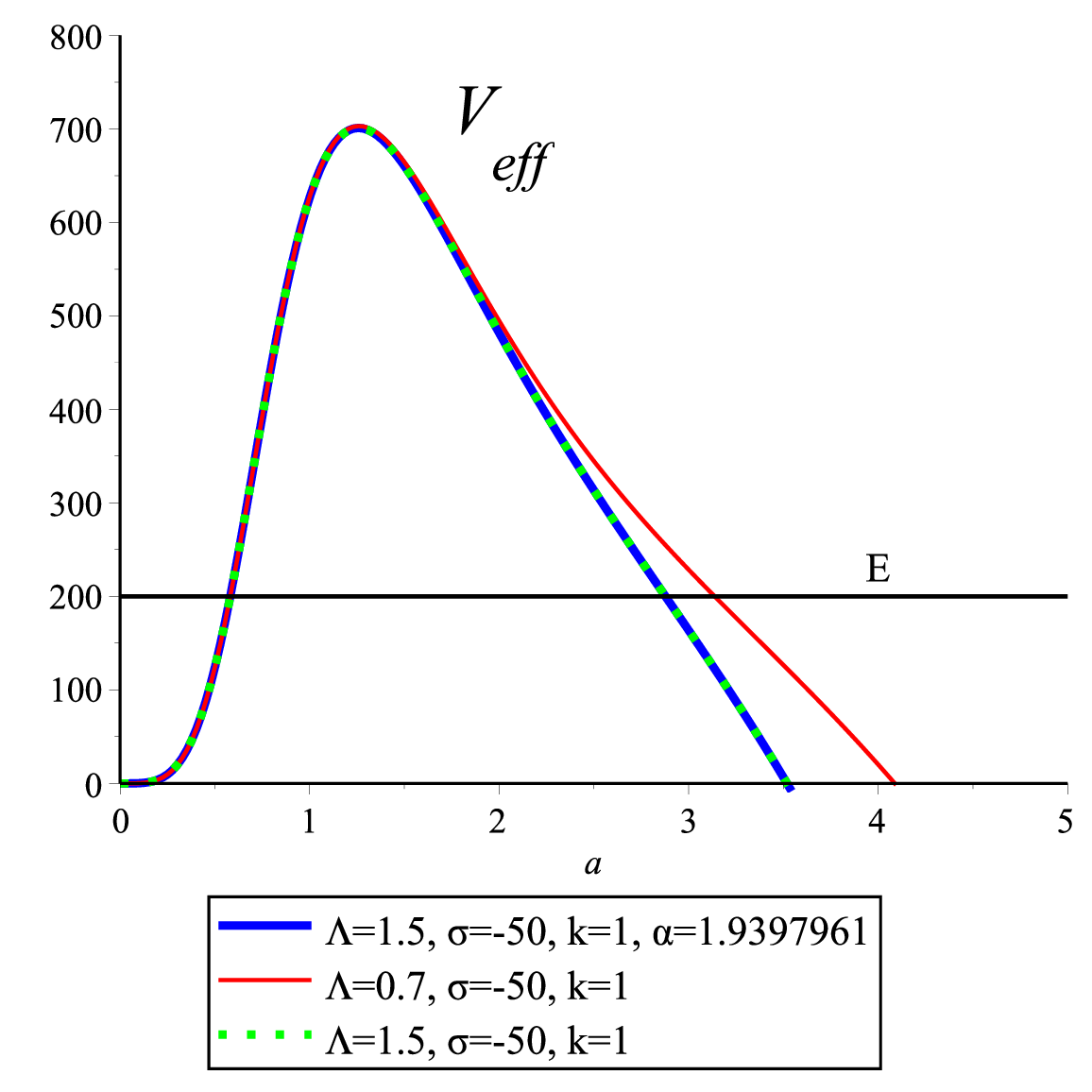}
  \caption{Effective potential for: fractional (blue line), $\Lambda=0.7$ (red line) and  \cite{GCM} model (green dot line), for $k=1$.}
  \label{V, k=1}
\end{figure}

\begin{figure}[H]
  \centering
  \includegraphics[width=0.55\textwidth]{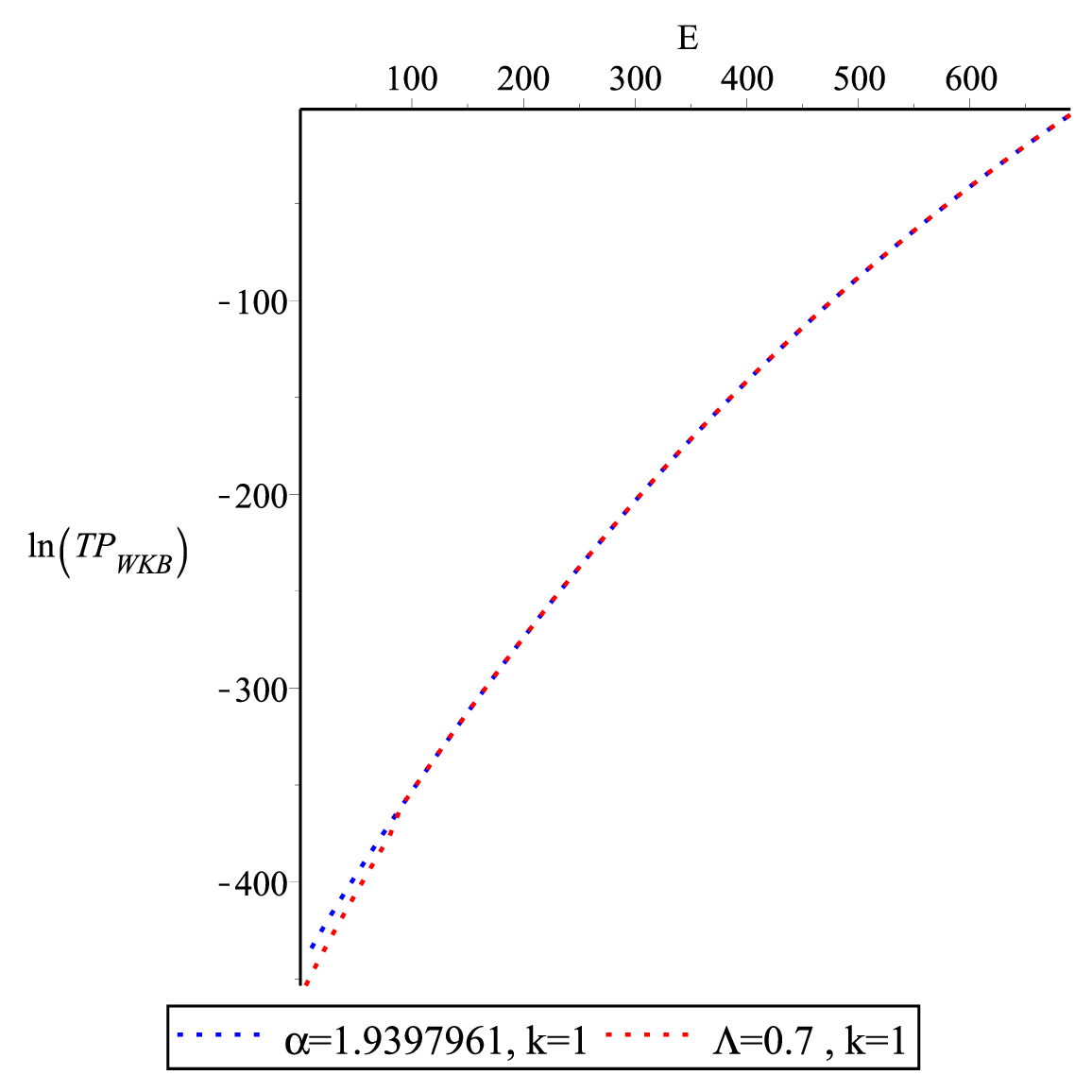}
  \caption{Comparison of $TP_{WKB}$, in logarithmic scale, for different energies $E$ in the fractional (blue slash line) and non-fractional $(\Lambda=0.7)$, red dot line, models for $k=1$.}
  \label{F não-F k=1}
\end{figure}

\FloatBarrier

By carefully contrasting the model in \cite{GCM} and the results herewith obtained from the model with fractional calculus, we find that the results of $TP_{WKB}$ for the non-fractional model with $\Lambda=0.7$ closely mimick the results found for the fractional model with $\Lambda=1.5$ and $\alpha = 1.9397961$, as we can see from the Figures \ref{F não-F k=-1}, \ref{F não-F k=0} and \ref{F não-F k=1}. The potential barrier of the non-fractional model with $\Lambda=0.7$ can be seen in Figures \ref{V, k=-1}-\ref{V, k=1}. \newline

This new and interesting result suggests that modifying the parameter  $\alpha$ in the fractional model may be comparable to modifying the dynamics of the potential barrier behavior in the non-fractional model, i.e., its height and width.
Using  $\theta$ as in  equation (\ref{theta}), which is used to appraise the height and width of the barrier $\cite{Merzbacher}$, we show in tables \ref{tabela 1}-\ref{tabela 6}  below some of the results for the same values of energy $E$ regarding the fractional and non-fractional model:

\begin{itemize}
    \item \underline{For hyperbolic space-like sections}
\end{itemize}

\begin{table}[h!]
\centering
\begin{tabular}{||c||c|c|c|}
\hline
$k$ & E & $\theta_{non-fractional}$  $(\Lambda=0.7)$ & $\theta_{fractional}$  $(\Lambda=1.5)$ and $\alpha = 1.9397961$ \\
\hline

\hline

%\multirow{5}{*}{-1} & 0 & $3.215640404\times 10^{90}$  & $3.215617252 \times 10^{90}$ \\

\multirow{5}{*}{-1} & 10 & $1.002778169\times 10^{88}$  & $1.303841382 \times 10^{88}$ \\

& 170 & $3.873356742\times 10^{58}$ & $1.577277123\times 10^{60}$ \\

& 340 & $3.066824521\times10^{34}$ & $1.353799484\times10^{36}$ \\

& 500 & $7.888358713\times10^{16}$ & $5.091334256\times10^{17}$ \\

& 660 & 463.1719831 & 400.5848243 \\
\hline
\end{tabular}
\caption{Comparative table of 
$\theta$ for different energies in the fractional and non-fractional models for $k=-1$.}
\label{tabela 1}
\end{table}

\begin{table}[h!]
\centering
\begin{tabular}{||c||c|c|c|}
\hline
$k$ & E & $TP_{WKB_{non-fractional}}$  $(\Lambda=0.7)$ & $TP_{WKB_{fractional}}$  $(\Lambda=1.5)$ and $\alpha = 1.9397961$ \\
\hline

\hline

%\multirow{5}{*}{-1} & 0 & $9.670858888\times 10^{-182}$  & $9.670998144 \times 10^{-182}$ \\

\multirow{5}{*}{-1} & 10 & $9.944667312\times 10^{-177}$  & $5.882344812 \times 10^{-177}$ \\

& 170 & $6.665381380\times 10^{-118}$ & $4.019610680\times 10^{-121}$ \\

& 340 & $1.063217573\times10^{-69}$ & $5.456212944\times10^{-73}$ \\

& 500 & $1.607040038\times10^{-34}$ & $3.857773981\times10^{-36}$ \\

& 660 & 0.000004661379104 & 0.000006231744824 \\
\hline
\end{tabular}
\caption{Comparative table of
tunneling probabilities for different energies in the fractional and non-fractional models for $k=-1$.}
\label{tabela 2}
\end{table}

\vspace{0.5cm}

\begin{itemize}
    \item \underline{For flat space-like sections}
\end{itemize}

\begin{table}[h!]
\centering
\begin{tabular}{||c||c|c|c|}
\hline
$k$ & E & $\theta_{non-fractional}$  $(\Lambda=0.7)$ & $\theta_{fractional}$  $(\Lambda=1.5)$ and $\alpha = 1.9397961$ \\
\hline

\hline
%\multirow{5}{*}{0} & 0 & $ 7.403548102\times10^{94}$  & $ 3.253577147\times10^{93} $ \\

\multirow{5}{*}{0} & 10 & $ 1.793016944\times10^{92}$  & $ 1.146483507\times10^{91} $ \\

&  170 & $1.379498224\times10^{61}$ & $1.674829609\times10^{62} $ \\

&  340 & $5.585366235\times10^{35}$ & $2.076589094\times10^{37}$ \\

&  500 & $3.790604212\times10^{17}$ & $2.537403708\times10^{18} $ \\

& 660 & 1191.527234  & 1069.113378 \\
\hline

\end{tabular}
\caption{Comparative table of 
the parameter 
$\theta$ for different energies in the fractional and non-fractional models for $k=0$.}
\label{tabela 3}
\end{table}

\vspace{0.5cm}

\begin{table}[h!]
\centering
\begin{tabular}{||c||c|c|c|}
\hline
$k$ & E & $TP_{WKB_{non-fractional}}$  $(\Lambda=0.7)$ & $TP_{WKB_{fractional}}$  $(\Lambda=1.5)$ and $\alpha = 1.9397961$ \\
\hline

\hline

%\multirow{5}{*}{0} & 0 & $1.824400554\times 10^{-190}$  & $9.446649068 \times 10^{-188}$ \\

\multirow{5}{*}{0} & 10 & $3.110507212\times 10^{-185}$  & $7.607892668 \times 10^{-183}$ \\

& 170 & $5.254818360\times 10^{-123}$ & $3.564993479\times 10^{-125}$ \\

& 340 & $3.205506704\times10^{-72}$ & $2.318989928\times10^{-75}$ \\

& 500 & $6.959581392\times10^{-36}$ & $1.553176674\times10^{-37}$ \\

& 660 & $6.959581392\times10^{-7}$ & $8.748876416\times10^{-7}$ \\
\hline
\end{tabular}
\caption{Comparative table of
%the parameter 
tunneling probabilities for different energies in the fractional and non-fractional models for $k=0$.}
\label{tabela 4}
\end{table}

\vspace{0.5cm}

\FloatBarrier

\begin{itemize}
    \item \underline{For spherical space-like sections}
\end{itemize}

\begin{table}[h!]
\centering
\begin{tabular}{||c||c|c|c|}
\hline
$k$ & E & $\theta_{non-fractional}$  $(\Lambda=0.7)$ & $\theta_{fractional}$  $(\Lambda=1.5)$ and $\alpha = 1.9397961$ \\
\hline

\hline
%\multirow{5}{*}{1} & 0 & $ 8.532079288\times 10^{99}$  & $ 6.252576211\times10^{96}$ \\

\multirow{5}{*}{1} & 10 & $ 1.577640978\times 10^{97}$  & $ 1.907005973\times10^{94}$ \\

 & 170 & $1.313683533\times 10^{64}$ & $ 2.830562481\times10^{64}$ \\
 & 340 & $1.380589165\times 10^{37}$ & $3.946000113\times 10^{38}$ \\
 & 500 & $1.992360778\times 10^{18}$ & $ 1.370930167\times10^{19}$ \\
 & 660 & 3160.340886 & 2945.792223 \\
\hline
\end{tabular}
\caption{Comparative table of
the parameter 
$\theta$ for different energies in the fractional and non-fractional models for $k=1$.}
\label{tabela 5}
\end{table}

\begin{table}[h!]
\centering
\begin{tabular}{||c||c|c|c|}
\hline
$k$ & E & $TP_{WKB_{non-fractional}}$  $(\Lambda=0.7)$ & $TP_{WKB_{fractional}}$  $(\Lambda=1.5)$ and $\alpha = 1.9397961$ \\
\hline

\hline

%\multirow{5}{*}{1} & 0 & 1.373694737$\times 10^{-200}$  & $2.557890873 \times 10^{-194}$ \\

\multirow{5}{*}{1} & 10 & $4.017756788\times 10^{-195}$  & $2.749766985 \times 10^{-189}$ \\

& 170 & $5.794533632\times 10^{-129}$ & $1.248114729\times 10^{-129}$ \\

& 340 & $5.246516932\times10^{-75}$ & $6.422229384\times10^{-78}$ \\

& 500 & $2.519208036\times10^{-37}$ & $5.320706876\times10^{-39}$ \\

& 660 & $1.001226000\times10^{-7}$ & $1.152380106\times10^{-7}$ \\
\hline
\end{tabular}
\caption{Comparative table of
tunneling probabilities for different energies in the fractional and non-fractional models for $k=1$.}
\label{tabela 6}
\end{table}

\vspace{0.5cm}

\FloatBarrier

Thus, analyzing the results qualitatively through the figures \ref{V, k=-1}-\ref{F não-F k=1} and quantitatively through the tables \ref{tabela 1}-\ref{tabela 6}, we see that the shape of the barrier and the results of the tunnelling probability of the fractional model with $\Lambda=1.5$ and $\alpha = 1.9397961$ are close to the shape of the barrier and to the results of the $TP_{WKB}$ for the non-fractional model with $\Lambda=0.7$, with the non-fractional parameters fixed equally. This convergence between the fractional and non-fractional models is lost for small $\Lambda$, on the order of $10^{-2}$, since the behaviour of the effective potential for $k=1$ is modified.

\section{Conclusion and Discussions} \label{conclusão}

In this work, we investigated whether a concrete tool of fractional calculus may influence the tunnelling probability via the $WKB$ approximation. For this purpose,  a $FRW$ model was employed, containing a radiation fluid, a positive cosmological constant, and an \textit{ad-hoc} potential. Moreover, we used the Riesz fractional derivative in the Hamiltonian operator, which was applied to the kinetic term. In this manner, we made it possible to investigate the behaviour of the tunnelling probability as a function of the new fractional parameter $\alpha$. \newline

Our results show that when $\alpha = 2$, we obtain the same results found in the study carried out in \cite{GCM}. 
When we allow $\alpha$ to vary, we observe that when we decrease $\alpha$ within its domain $1 < \alpha \leq 2$, we found that the probability of tunnelling decreases. Thus, the Universe is more likely to be created for higher values of $\alpha$, namely $\alpha=2$.\newline

These new results also 
 allowed us to suggest that although our potential barrier is the same as that established in \cite{GCM}, applying fractional calculus in the kinetic term of the Hamiltonian operator modified the tunnelling probability. Upon investigating this feature in more detail, we identified that the decrease of the parameter $\alpha$
 seems to produce what could have been caused had we instead modified the non-fractional setup's barrier in \cite{GCM}. As if making it broader and higher through modifying its sole parameters $\Lambda, \sigma, k$ (where no fractional elements are present). To substantiate this argument, we choose suitable values for the tunnelling probability in the fractional model and the non-fractional model, and we could find that, e.g.,  taking  $\Lambda=1.5$ and $\alpha = 1.9397961$ in the fractional setting allowed to reproduced very closely the probability computed for the non-fractional model with $\Lambda=0.7$. This was checked for other ranges, as presented in the previous section.  Please compare tables \ref{tabela 1} to \ref{tabela 6} and figures \ref{V, k=-1}-\ref{F não-F k=1}. \newline

In addition to the qualitative analysis for the potential barriers in the fractional and non-fractional models, through the figures \ref{V, k=-1}-\ref{F não-F k=1}, we performed a quantitative analysis through  $\theta$, which informs about the height and length of the barrier. The results found in tables \ref{tabela 1}-\ref{tabela 3} confirm the similarity of the barriers and the results of the tunnelling probability between the models, namely the fractional contrasted with the non-fractional picture. \newline

Finally, our results also allow for an interesting discussion between selecting values for $\Lambda$ (in a non-fractional conventional set-up) versus balancing, e.g., both $\Lambda$ and $\alpha$ in the fractional framework. Concretely, as pointed out above, the probability transition in the former if, e.g., $\Lambda=0.7$,  is very close to the value computed if in the latter we employ instead, e.g., $\Lambda=1.5, \alpha=1.9397961$. This suggests that if a fractional element could be present, a universe with a larger $\Lambda$ could tunnel through, mimicking the tunneling of another universe with  a smaller value of $\Lambda$ and instead in a non-fractional setting. \newline

These new results also contribute and shed additional light on the use of fractional calculus in quantum cosmology, more precisely on the probability of tunnelling, by using a more generic potential than those studied in \cite{Hasan} and \cite{Oliveira}. Lastly, the application of fractional calculus in quantum cosmology, more precisely in calculating the tunnelling probability, is relatively new and requires further investigation.

\section*{Acknowledgements}
 D. L. Canedo thanks Coordenação de Aperfeiçoamento de Pessoal de Nível Superior (CAPES) and Universidade Federal de Juiz de Fora (UFJF) for his scholarships and the Universidade da Beira Interior (UBI) for academic support in the PDSE-CAPES program. PM acknowledges the FCT grant UID-B-MAT/00212/2020 at CMA-UBI plus
the COST Actions CA23130 (Bridging high and low energies in search of
quantum gravity (BridgeQG)) and CA23115 (Relativistic Quantum Information (RQI)). G. Oliveira-Neto thanks FAPEMIG (APQ-06640-24) for partial financial support.

\appendix

\renewcommand{\theequation}{A.\arabic{equation}} % Define numeração das equações como A.1, A.2, ...
\setcounter{equation}{0} % Reinicia a numeração das equações

\section*{Appendix: $D_{\alpha}$  Analysis}\label{Constante} % Apêndice sem numeração

In section \ref{seção 3.1}, we reported some arbitrariness in choosing the constant $D_{\alpha}$. This constant is a function of the Riesz parameter $\alpha$, which also appears explicitly in the tunnelling probability expression (\ref{PTwkb3}). For simplicity, we will rewrite this equation here:
\begin{equation*} 
     TP_{WKB} \equiv \frac{4}{\Big{(}2e^{\int_{a_{L}}^{a_{R}} \kappa(a) da} + \frac{1}{2e^{\int_{a_{L}}^{a_{R}} \kappa(a) da}}\Big{)}^{2}}
\end{equation*}
where $\kappa(a) = \Bigg{(}{\frac{1}{D_{\alpha}}  \Big{(}3ka^{2} - \Lambda a^{4} + \frac{\sigma^{2} a^{4}}{(a^{3} + 1)^{2}} - E\Big{)}}\Bigg{)}^{\frac{1}{\alpha}}$. \newline

By analyzing this equation in detail, we can see that the fractional exponent $\alpha$ present in the term $\kappa(a)$ acts faster than the constant $D_{\alpha}$, a multiplicative factor. Thus, the definition of $D_{\alpha}$ only changes the result quantitatively, not qualitatively.\newline

 To determine the influence of the constant $D_{\alpha}$ on the tunnelling probability, we present below an analysis for two choices of this constant: $D_{\alpha} = (\frac{1}{12})^{\frac{\alpha}{2}}$ and $D_{\alpha} = (\frac{1}{12})^{\frac{2}{\alpha}}$. Note that for $\alpha=2$, both choices result in $D_{\alpha} = \frac{1}{12}$.\newline

  In the figures below, we can see the probability of tunnelling, in logarithmic scale, for different energies $E$, different curvatures $k$ and two values of the fractional parameter $\alpha$, for example. We have the parameters $\Lambda = 1.5$ and $\sigma=-50$ fixed in both figures.
\begin{figure}[H]
  \centering
  \includegraphics[width=0.55\textwidth]{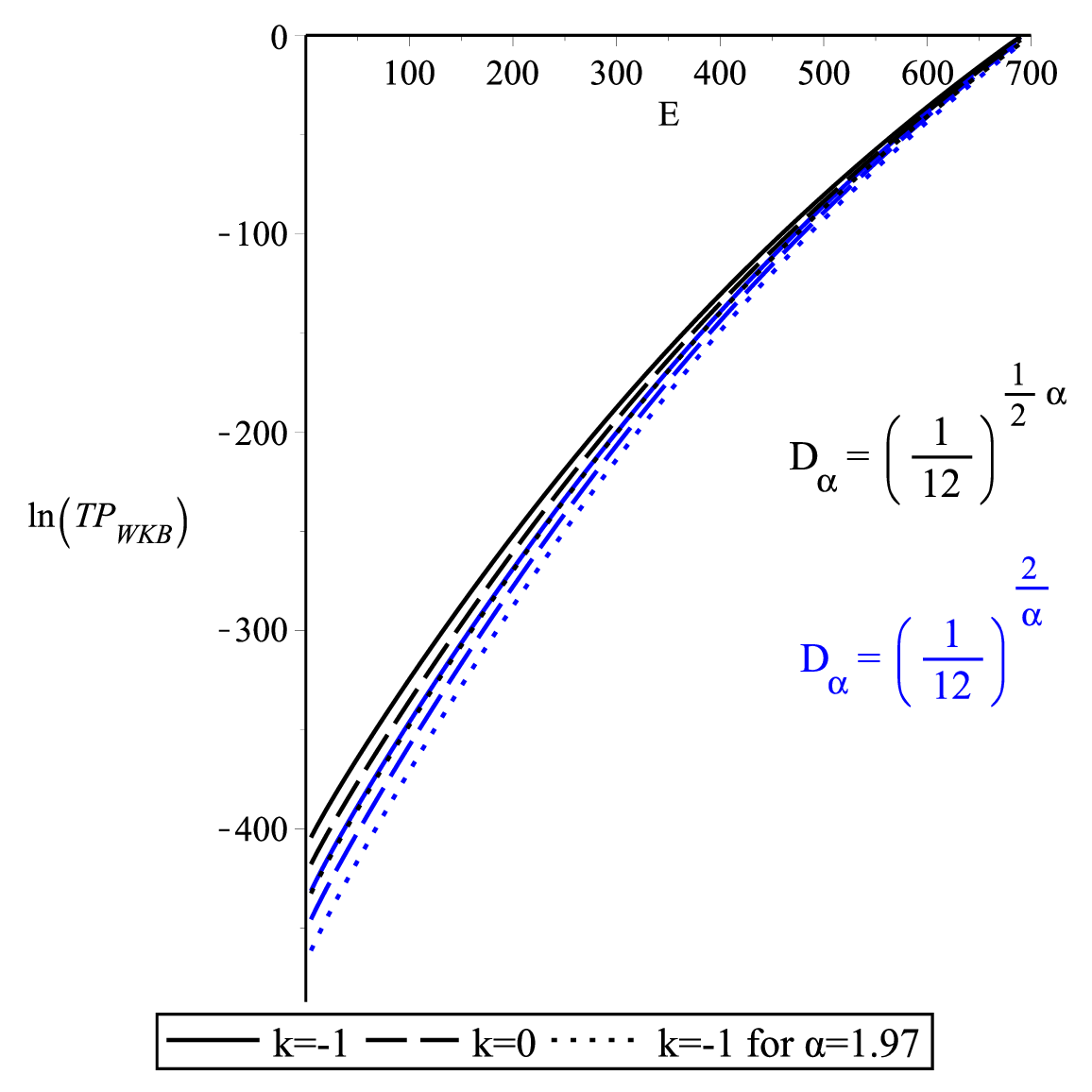}
  \caption{Behavior of tunnelling probability, in logarithmic scale, as a function of energy $E$ and curvature $k$, when $\alpha=1.97$, $\Lambda = 1.5$ and $\sigma=-50$, for $D_{\alpha} = (\frac{1}{12})^{\frac{\alpha}{2}}$ and $D_{\alpha} = (\frac{1}{12})^{\frac{2}{\alpha}}$. }
  \label{d_a,a=1.97}
\end{figure}

\begin{figure}[H]
  \centering
  \includegraphics[width=0.55\textwidth]{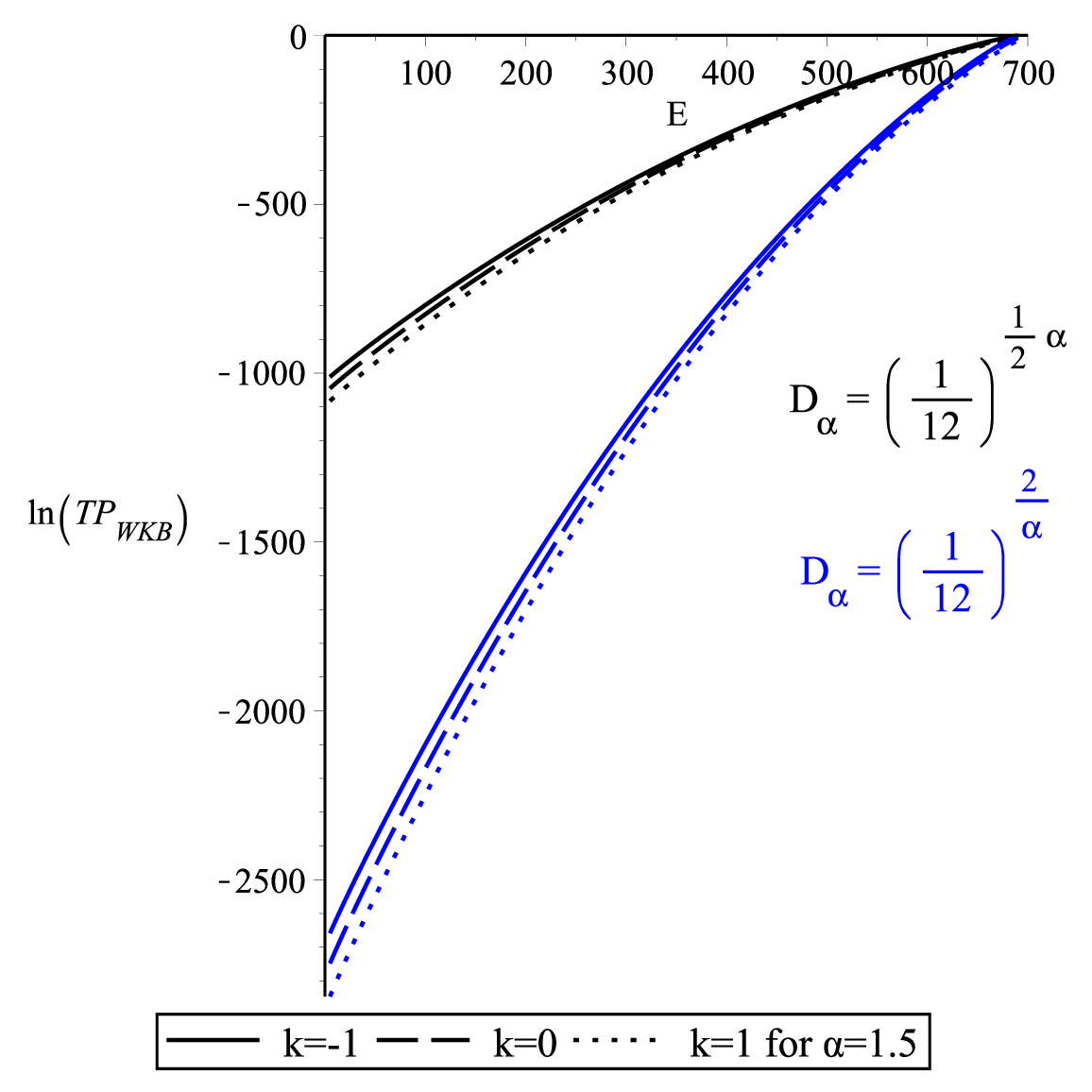}
  \caption{Behavior of tunnelling probability, in logarithmic scale, as a function of energy $E$ and curvature $k$, when $\alpha=1.5$, $\Lambda = 1.5$ and $\sigma=-50$, for $D_{\alpha} = (\frac{1}{12})^{\frac{\alpha}{2}}$ and $D_{\alpha} = (\frac{1}{12})^{\frac{2}{\alpha}}$. }
  \label{d_a,a=1.5}
\end{figure}

\noindent Thus, it can be observed that the constant $D_{\alpha}$ can be more influential in the result of the tunnelling probability when $\alpha \to 1$, where $\alpha \in (1,2]$. However, as discussed at the beginning of the appendix, the behaviour of the tunnelling probability as a function of the model parameters remains unchanged.

\end{document}